# THE PROBABILITY OF A ROBUST INFERENCE FOR INTERNAL VALIDITY AND ITS APPLICATIONS IN REGRESSION MODELS

**Tenglong Li**

**Department of Biostatistics, Boston University**

**Email: litenglo@bu.edu**

**Kenneth A. Frank**

**Department of Counseling, Educational Psychology and Special Education, Michigan State University**

**Email: kenfrank@msu.edu**

**Abstract**

The internal validity of observational study is often subject to debate. In this study, we define the unobserved sample based on the counterfactuals and formalize its relationship with the null hypothesis statistical testing (NHST) for regression models. The probability of a robust inference for internal validity, i.e., the PIV, is the probability of rejecting the null hypothesis again based on the ideal sample which is defined as the combination of the observed and unobserved samples, provided the same null hypothesis has already been rejected for the observed sample. When the unconfoundedness assumption is dubious, one can bound the PIV of an inference based on bounded belief about the mean counterfactual outcomes, which is often needed in this case. Essentially, the PIV is statistical power of the NHST that is thought to be built on the ideal sample. We summarize the process of evaluating internal validity with the PIV into a six-step procedure and illustrate it with an empirical example (i.e., Hong and Raudenbush (2005)).

# 1-Introduction

Causal inferences are often made from observational studies based on the combination of regression modelling and propensity score techniques (Gelman and Hill, 2007; Murnane and Willett, 2011; Imbens and Rubin, 2015; Morgan and Winship, 2015). Internal validity, which refers to whether one can infer a causal relationship between two variables given they are correlated (Shadish et al. 2002), is difficult to evaluate and often in doubt since there is no randomization involved in making a causal inference in observational study (Rosenbaum and Rubin, 1983b; Rosenbaum, 2002, 2010; Imai et al. 2008). Causal inference is essentially a missing data problem based on the key concept of potential outcomes which refers to the outcomes under all possible treatments for each subject (Holland, 1986; Rubin, 2008): Regardless of the treatment assignment, only one of the potential outcomes can be realized and the others are missing for every individual (the missing potential outcomes are called the counterfactual outcomes) (Rubin, 2005, 2007, 2008; Imbens and Rubin, 2015). Most inferences in observational studies assume "unconfoundedness", which states that the counterfactual outcomes would be missing at random (MAR) conditional on a set of covariates, and thus suggests internal validity should not be compromised by the lack of randomization once the pivotal covariates are controlled (Rosenbaum and Rubin, 1983a; Imbens, 2004).

Because the unconfoundedness assumption is hardly testable (Heckman, 2005; Rosenbaum and Rubin 1983b; Rosenbaum, 1987), one may suspect the counterfactual outcomes are not MAR and thus a causal inference may be invalidated. The robustness of a causal inference is defined in this context as whether a causal relationship between two variables can still hold when the unconfoundedness assumption fails. To evaluate the robustness of a causal inference, a belief about the counterfactual outcomes or missing confounders is typically required so that one could decide whether an inference is still valid based on such belief (Frank, 2000; Frank et al. 2013). Our goal in this paper is to quantify the robustness of a causal inference in observational study based on one's belief about the mean counterfactual outcomes for the treated subjects and the controlled subjects (Frank and Min 2007; Frank et al. 2013). This paper is organized as follows: we first define the counterfactual data and the unobserved sample, both of which are built on the counterfactual outcomes. Next, we define the ideal sample as the combination of the observed and unobserved samples, which as the name suggests is ideal for making causal inference (Sobel, 1996; Rubin, 2004, 2005; Frank et al. 2013). The probability of a causal inference is robust for internal validity (henceforth abbreviated as the PIV) is defined based on the ideal sample and served as the robustness index of a causal inference. The robustness of a causal inference is informed by the bound(s) of the PIV which can be obtained based on belief about the mean counterfactual outcomes. To illustrate this approach, we quantify the robustness of the inference of Hong & Raudenbush (2005) which found a significant negative effect of kindergarten retention on reading achievement. The inference of Hong & Raudenbush (2005) was built on a nationally representative sample and a design based on propensity score stratification, given the treatments (retained in kindergarten vs promoted to the first grade) were impossible to be randomly assigned to students, particularly raising concerns about its internal validity (Schafer and Kang, 2008; Allen et al. 2009; Hong, 2010; Frank et al., 2013).

**2-The unobserved sample and the ideal sample**

**2.1-Research setting**

Throughout this paper, we assume a causal inference has been made in an observational study with two groups (i.e., the treatment group and the control group) and a representative sample such that its internal validity is the major concern. In this paper, the regression estimator of average treatment effect refers to the estimated beta coefficient[1] of the treatment indicator W, i.e., $\hat{\beta}_W$ in the regression $Y = \beta_0 + \beta_W W + \beta_1 Z_1 + \cdots + \beta_p Z_p + \varepsilon$, where W=1 for treatment cases, 0 for the control. The covariates included are $Z_1, Z_2, \ldots, Z_p$, which are usually a subset of the covariates that need to be controlled for the unconfoundedness assumption to hold. Such regression model is often built on propensity score analysis, which means the estimated propensity scores and/or the propensity score design (propensity score matched pairs or strata) need to be controlled in the regression as well.

**2.2-Definitions**

**Definition 1**: **The counterfactual data** for a subject refers to the imaginary observation where the outcome is the counterfactual outcome, the treatment is different than what is observed and the covariates' values are identical to the ones in his observed data. Thus, there are no variables confounded with treatment assignment when both the counterfactual data and the observed data are included (Holland, 1986).

**Example:** In Hong & Raudenbush (2005), the counterfactual data for John who was retained in the kindergarten would be John's potential reading score had he been promoted to first grade, and the covariates (e.g., gender, race, socioeconomic status) were identical to those in his observation.

**Definition 2**: **The unobserved sample** refers to the counterfactual data for all sampled subjects.

**Example:** The unobserved sample of Hong & Raudenbush (2005) is the counterfactual data for all the sample students in their study.

---

[1] We define $\beta_W$ as the beta coefficient in order to standardize the discussion, but one can always apply our framework to an ordinary regression coefficient of the treatment indicator with the necessary transformation.

Figure 1 illustrates the conceptualization of the unobserved sample in Hong & Raudenbush (2005) for the regression estimator. Let $Y_{r,i}^{ob}$ and $Y_{p,j}^{ob}$ be the observed reading scores for the retained students and the promoted students respectively, and their corresponding counterfactual reading scores are denoted by $Y_{p,i}^{un}$ and $Y_{r,j}^{un}$. $R_i$ denotes the observed data for any student i who was retained in the kindergarten and is written as $[Y_{r,i}^{ob}, W=1, Z_{1,i}, Z_{2,i}, \ldots, Z_{p,i}]$, and the corresponding counterfactual data $P_i$ should be $[Y_{p,i}^{un}, W=0, Z_{1,i}, Z_{2,i}, \ldots, Z_{p,i}]$. Likewise, $P_j$ denotes the observed data of any student j who was promoted to the first grade and is written as $[Y_{p,j}^{ob}, W=0, Z_{1,j}, Z_{2,j}, \ldots, Z_{p,j}]$ and the corresponding counterfactual data $R_j$ is $[Y_{r,j}^{un}, W=1, Z_{1,j}, Z_{2,j}, \ldots, Z_{p,j}]$. By our definition, the unobserved sample consists of the counterfactual data $P_i$ and $R_j$.

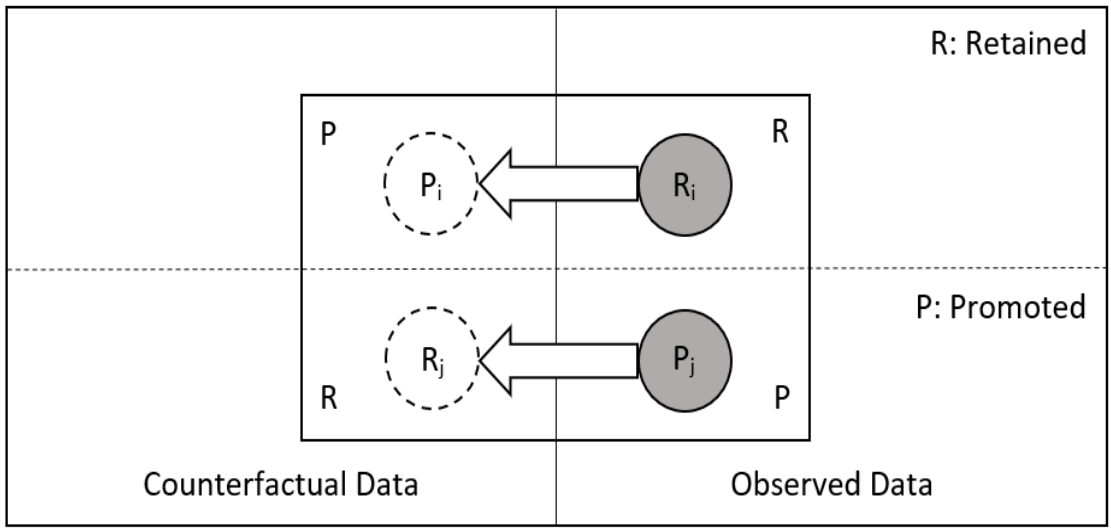

*Figure 1.* Observed and counterfactual data for kindergarten retention for the regression estimator

**Definition 3**: **The ideal sample** refers to the combination of the observed data and the unobserved data for all sampled subjects. We argue that causal inference should be based on the ideal sample rather than the observed sample. When the unconfoundedness assumption fails, the unobserved sample is distinct from the observed sample, implying a gap between the observed sample and the ideal sample. Essentially, the robustness of a causal inference hinges on the differences between the unobserved sample and the observed sample. Therefore, the central task in evaluating the robustness of an inference should be examining the relationship between the unobserved sample and an inference based on the ideal sample.

**2.3-Notation**

This section introduces the notation for the parameter and the sample statistics. The parameter refers to the average treatment effect for the target population, which is denoted by $\beta_W$, i.e., the beta coefficient of W in the regression model. The PIV is built on the distribution of $\beta_W$, whose randomness is mainly due to the counterfactual outcomes and the fact that they might be incomparable to the observed outcomes considering the potential failure of the unconfoundedness assumption. Therefore, the distribution of $\beta_W$ needs to be defined based on the ideal sample in order to take all potential outcomes into consideration. For this purpose, we need to define notation for the sample statistics for the observed, unobserved and ideal samples separately.

The observed sample statistics (known and fixed): $\hat{\beta}_W$ denotes the estimated beta coefficient of W based on the observed sample. $\bar{Y}_t^{ob}$ denotes the adjusted mean outcome of the observed treated subjects and $\bar{Y}_c^{ob}$ denotes the adjusted mean outcome of the observed control subjects, adjusted based on the regression model described earlier. $\hat{\sigma}_t^2$ and $\hat{\sigma}_c^2$ denote the variances of the treated and the control outcomes in the observed sample respectively. $n^{ob}$ denotes the observed sample size and $\pi$ denotes the proportion of treated subjects in the observed sample. $R^2$ denotes the estimated R-square for the regression model.

The unobserved sample statistics (focused unknown): $\bar{Y}_t^{un}$ and $\bar{Y}_c^{un}$ are the mean counterfactual outcomes for the treated subjects and the control subjects respectively. We will show the PIV is a function of them, conditional on the observed sample statistics.

The ideal sample statistics (unknown due to the unobserved sample): $\hat{\beta}_W^{id}$ is the regression estimator of average treatment effect based on the ideal sample and $se(\hat{\beta}_W^{id})$ is its standard error. $\bar{Y}_t^{id}$ and $\bar{Y}_c^{id}$ denote the mean outcomes for the treated subjects and the control subjects in the ideal sample respectively.

**3-The probability of a causal inference is robust for internal validity**

The PIV is rooted in the context of null hypothesis statistical testing (NHST). To decide whether there is an effect, the null hypothesis $H_0$: $\beta_W = 0$ is tested against the alternative hypothesis $H_a$: $\beta_W \neq 0$.[2] Here $\beta_W^{\#}$ is the threshold for a significant finding (rejecting the null hypothesis) regarding the beta coefficient $\beta_W$, and for NHST $\beta_W^{\#}$ is just the product of a critical value C and the standard error of $\hat{\beta}_W$. The PIV is meaningful only if the null hypothesis has been rejected based on the observed sample. For example, Hong and Raudenbush (2005) rejected the null hypothesis of zero effect of kindergarten retention on achievement based on their observed sample. Since the counterfactual outcomes might be distinct from the observed outcomes, it's natural to wonder whether the null hypothesis would be rejected again based on the ideal sample if the counterfactual outcomes were known.

Frank et al. (2013) proposed a set of decision rules when a causal inference would be invalidated due to limited internal validity, and they can be adapted to regression models as follows: Given $\hat{\beta}_W$ is significantly positive, an inference would be invalidated if $\hat{\beta}_W > \beta_W^\# > \beta_W$. Given $\hat{\beta}_W$ is significantly negative, an inference would be invalidated if $\hat{\beta}_W < \beta_W^\# < \beta_W$. The decision rules can be also interpreted in the opposite way: an inference would be valid if $\beta_W > \beta_W^\#$ for an observed significant positive estimate or $\beta_W < \beta_W^\#$ for an observed significant negative estimate, given $\hat{\beta}_W$ is fixed and exceeds the threshold $\beta_W^\#$ (so that it is significant). Drawing on this interpretation, the probability of a causal inference is robust for internal validity (PIV) is defined as follows for an observed significant positive estimate, for the ideal sample $\mathbf{D^{id}}$ (consisting of the observed and unobserved samples):

$$P(\beta_W > \beta_W^\# \mid \mathbf{D^{id}}) \qquad (1)$$

---

[2] Our framework should be easily modified for constants other than 0 or one-sided hypothesis.

Likewise, the PIV is defined as follows for an observed significant negative estimate:

$$P(\beta_W < \beta_W^\# \mid \mathbf{D^{id}}) \qquad (2)$$

The PIV defined above can be best understood as a probability defined based on the posterior distribution of $\beta_W$ where its prior distribution is thought to be built on the unobserved sample while the likelihood function is built on the observed sample. Interestingly, the posterior distribution in this case can be thought as a distribution that is built on the ideal sample (Li, 2018). Under Bayesian framework, $\beta_W$ shall be treated as random and the threshold $\beta_W^\#$ shall be treated as a fixed (but unknown) quantity defined by the ideal sample. The posterior distribution of $\beta_W$ is detailed in the next section.

It is noteworthy that the PIV in (1) and (2) are simplified from $P(\beta_W > \beta_W^\# \mid \hat{\beta}_W > \beta_W^\#, \mathbf{D^{id}})$ and $P(\beta_W < \beta_W^\# \mid \hat{\beta}_W < \beta_W^\#, \mathbf{D^{id}})$ respectively. The condition $\hat{\beta}_W > \beta_W^\#$ (or $\hat{\beta}_W < \beta_W^\#$) is redundant as it is conveyed by inference from the observed sample, which is a part of $\mathbf{D^{id}}$. The PIV essentially is the probability of rejecting the null hypothesis again for the ideal sample given the same null hypothesis has been rejected for the observed sample, when the counterfactual data has been included. By definition, the PIV is the statistical power of retesting the null hypothesis: $\beta_W = 0$ versus the alternative hypothesis: $\beta_W = \hat{\beta}_W^{id}$ ($\hat{\beta}_W^{id} \neq 0$) based on the ideal sample. For a NHST that is built on either normal or student T distribution, the PIV has the following relationship with the T-ratio $T = \dfrac{\hat{\beta}_W^{id}}{se(\hat{\beta}_W^{id})}$:

For a significantly positive $\hat{\beta}_W$ and a critical value C, we have:

$$probit(PIV) = T - C \qquad (3)$$

For a significantly negative $\hat{\beta}_W$ and a critical value C, we have:

$$probit(PIV) = C - T \qquad (4)$$

We caution readers that (3) and (4) are approximately true for studies with small sample sizes and C corresponds to the chosen level of significance. For example, C would be 1.96 if $\hat{\beta}_W^{id}$ is significantly positive and the level of significance is 0.05.

**4-The relationship between the PIV and the unobserved sample**

**Theorem 1:** Given the unobserved sample and the R square for regression, the distribution of $\beta_W$ conditional on the ideal sample is as follows based on a classical linear regression model (CLRM):

$$\beta_W \mid \mathbf{D^{id}} \sim N(\dfrac{\bar{Y}_t^{id} - \bar{Y}_c^{id}}{\sqrt{2\hat{\sigma}_t^2 + 2\pi(1-\pi)[(\bar{Y}_t^{un} - \bar{Y}_t^{ob})^2 + (\bar{Y}_c^{un} - \bar{Y}_c^{ob})^2] + 2\hat{\sigma}_c^2 + (\bar{Y}_t^{id} - \bar{Y}_c^{id})^2}}, \dfrac{1-R^2}{2n^{ob}}) \qquad (3)$$

where:

$$\bar{Y}_t^{id} = (1-\pi)\bar{Y}_t^{un} + \pi\bar{Y}_t^{ob}$$
$$\bar{Y}_c^{id} = \pi\bar{Y}_c^{un} + (1-\pi)\bar{Y}_c^{ob} \qquad (4)$$

It is clear that, the distribution of $\beta_W$ is conditional on the values of mean counterfactual outcomes $\bar{Y}_t^{un}$ and $\bar{Y}_c^{un}$ besides the observed sample statistics, so they need to be conceptualized. $\bar{Y}_t^{id}$ is weighted average of $\bar{Y}_t^{un}$ and $\bar{Y}_t^{ob}$, with the weight defined by $\pi$. For the example of the effect of kindergarten retention on reading achievement (Hong and Raudenbush, 2005), $\bar{Y}_t^{un}$ is the mean reading score for the promoted students had they all been retained instead and $\bar{Y}_t^{ob}$ is the observed mean reading score for the retained students, with the weight defined by the proportion of students who were retained in the observed sample. Likewise, $\bar{Y}_c^{id}$ is weighted average of the mean reading score for the retained students had they all been promoted instead ($\bar{Y}_c^{un}$) and the observed mean reading score for the promoted students ($\bar{Y}_c^{ob}$).

Under the Bayesian framework, $\bar{Y}_t^{un}$ and $\bar{Y}_c^{un}$ characterize the unobserved sample which defines the prior distribution of $\beta_W$, as literature suggests that prior can be treated as a function of the data of particular interest (Diaconis and Ylvisaker, 1979, 1985; Frank and Min, 2007; Hoff, 2009; Pearl and Mackenzie, 2018). In this case, different values of $\bar{Y}_t^{un}$ and $\bar{Y}_c^{un}$ represent different prior beliefs about $\beta_W$ and shape the posterior distribution of $\beta_W$, which is shown to be same as the distribution of $\beta_W$ built on the ideal sample (Li, 2018; Li and Frank, 2019, 2020).

The distribution of $\beta_W$ only depends on $\bar{Y}_t^{un}$ and $\bar{Y}_c^{un}$, assuming the observed sample statistics are known and fixed. As a result, the relationship between the PIV and the two mean counterfactual outcomes is derived from theorem 1 as follows:

**Theorem 2:** Built on theorem 1, the probit link of PIV is a function of $\bar{Y}_t^{un}$ and $\bar{Y}_c^{un}$, conditional on the observed sample statistics $R^2, n^{ob}, \bar{Y}_t^{ob}, \bar{Y}_c^{ob}, \hat{\sigma}_t^2, \hat{\sigma}_c^2, \pi$ as well as the threshold $\beta_W^{\#}$ for rejecting the null hypothesis. Specifically, for a significant positive $\hat{\beta}_W$, we have:

$$probit(PIV) = \frac{\sqrt{2n^{ob}}}{\sqrt{1-R^2}}[\frac{\bar{Y}_t^{id} - \bar{Y}_c^{id}}{\sqrt{2\hat{\sigma}_t^2 + 2\pi(1-\pi)[(\bar{Y}_t^{un} - \bar{Y}_t^{ob})^2 + (\bar{Y}_c^{un} - \bar{Y}_c^{ob})^2] + 2\hat{\sigma}_c^2 + (\bar{Y}_t^{id} - \bar{Y}_c^{id})^2}} - \beta_W^{\#}] \quad (5)$$

For a significant negative $\hat{\beta}_W$, we have:

$$probit(PIV) = \frac{\sqrt{2n^{ob}}}{\sqrt{1-R^2}}[\beta_W^{\#} - \frac{\bar{Y}_t^{id} - \bar{Y}_c^{id}}{\sqrt{2\hat{\sigma}_t^2 + 2\pi(1-\pi)[(\bar{Y}_t^{un} - \bar{Y}_t^{ob})^2 + (\bar{Y}_c^{un} - \bar{Y}_c^{ob})^2] + 2\hat{\sigma}_c^2 + (\bar{Y}_t^{id} - \bar{Y}_c^{id})^2}}] \quad (6)$$

where $\bar{Y}_t^{id}$ and $\bar{Y}_c^{id}$ are provided in (4).

The decision threshold $\beta_W^{\#}$ could be either a fixed value that is pragmatically set based on transaction cost/policy implication/literature review (Frank et al., 2013) or a statistical threshold that is a product between the critical value and the standard error, i.e., $C * se(\hat{\beta}_W^{id})$, where the critical value C is chosen based on the level of significance. $se(\hat{\beta}_W^{id})$ is computed based on R-square in the regression and the observed sample size as follows:

$$se(\hat{\beta}_W^{id}) = \sqrt{\frac{1-R^2}{2n^{ob}}} \quad (7)$$

Plugging $\beta_W^{\#}$ as $C * se(\hat{\beta}_W^{id})$ into (5), the relationship between PIV and the unobserved sample for a significant positive $\hat{\beta}_W$ yields:

$$probit(PIV) = \frac{\sqrt{2n^{ob}}}{\sqrt{1-R^2}} \frac{\bar{Y}_t^{id} - \bar{Y}_c^{id}}{\sqrt{2\hat{\sigma}_t^2 + 2\pi(1-\pi)[(\bar{Y}_t^{un} - \bar{Y}_t^{ob})^2 + (\bar{Y}_c^{un} - \bar{Y}_c^{ob})^2] + 2\hat{\sigma}_c^2 + (\bar{Y}_t^{id} - \bar{Y}_c^{id})^2}} - C \quad (8)$$

Likewise, by plugging $\beta_W^{\#}$ as $C*se(\hat{\beta}_W^{id})$ into (6), the relationship between PIV and the unobserved sample for a significant negative $\hat{\beta}_W$ yields:

$$probit(PIV) = C - \frac{\sqrt{2n^{ob}}}{\sqrt{1-R^2}} \frac{\bar{Y}_t^{id} - \bar{Y}_c^{id}}{\sqrt{2\hat{\sigma}_t^2 + 2\pi(1-\pi)[(\bar{Y}_t^{un} - \bar{Y}_t^{ob})^2 + (\bar{Y}_c^{un} - \bar{Y}_c^{ob})^2] + 2\hat{\sigma}_c^2 + (\bar{Y}_t^{id} - \bar{Y}_c^{id})^2}} \quad (9)$$

The aforementioned relationships between the PIV and the unobserved sample allow one to bound the PIV based on his/her (bounded) belief about $\bar{Y}_t^{un}$ and $\bar{Y}_c^{un}$. For the example of Hong & Raudenbush (2005), if one believes the mean reading score of the promoted students had they been retained instead (i.e., $\bar{Y}_t^{un}$) is no larger than their mean observed reading score (45.78) and the mean reading score of the retained students had they been promoted instead (i.e., $\bar{Y}_c^{un}$) equals the grand mean (45.2), the lower bound of the PIV would be 0.92. However, if his belief is modified as $\bar{Y}_t^{un} \leq 45.78$ and $\bar{Y}_c^{un} \geq 44$, the lower bound of the PIV would be 0.82 instead.

## 5-Example: The effect of kindergarten retention on reading achievement

### 5.1-Overview

Kindergarten retention is estimated to affect 7 percent to 15 percent of the student population in the U.S. and cost $20 billion dollars annually (Alexander et al., 2003; Tingle et al., 2012). It also imposes physical and psychological costs on retained students, and thus has been a controversial issue for many years. To examine the effectiveness of kindergarten retention, Hong & Raudenbush (2005) conducted propensity score analysis using nationally representative data from the Early Childhood Longitudinal Study (ECLS) and a rich set of covariates such as student background information, psychological/motivational measures as well as pretests. Based on a multilevel model which controlled for both the logit of propensity scores as well as the propensity score strata, they estimated the effect of kindergarten retention on students' reading

achievement as -9.01 with standard error of 0.68, which amounted to an effect size of 0.67. Ultimately, Hong and Raudenbush concluded that retention reduces achievement: "children who were retained would have learned more had they been promoted (page 200)."

However, the internal validity of Hong & Raudenbush (2005) is open to debate since it hinges on the unconfoundedness assumption, which forbids any potential confounding variable to be left unobserved and uncontrolled in the causal model. Nonetheless, Frank et al. (2013) has argued that some potential confounders, such as key measures of cognitive ability and emotional disposition, might still be missing in their propensity score model and thus introduce selection bias in the estimate. If an omitted confounder were negatively correlated with kindergarten retention and positively correlated with reading achievement, the estimate of kindergarten retention could be downwardly biased, and thus their inference would be invalidated if such omitted confounder were taken into account.

To quantify the concern so that researchers and policymakers may evaluate to robustness of the inference to omitted confounders, we develop an analytical procedure based on the relationship between the PIV and the mean counterfactual outcomes in theorem 2. Specifically, this analytical procedure has six steps: 1-get the observed sample statistics, 2-choose critical value $C^3$, 3-obtain the relationship between the PIV and the mean counterfactual outcomes, 4-state belief about the mean counterfactual outcomes, 5-bound the PIV, 6-conclusion.

---

[3] Here we assume one use a statistical threshold, but a decision threshold could be a non-statistical one.

## 5.2-Quantifying the robustness of the inference of Hong & Raudenbush (2005)

1-Get the observed sample statistics: The observed sample statistics $R^2, n^{ob}, \bar{Y}_t^{ob}, \bar{Y}_c^{ob}, \hat{\sigma}_t^2, \hat{\sigma}_c^2, \pi$ are obtained as follows: $R^2 = 0.36, n^{ob} = 7639, \bar{Y}_t^{ob} = 36.77, \bar{Y}_c^{ob} = 45.78, \hat{\sigma}_t^2 = 143.26, \hat{\sigma}_c^2 = 138.83, \pi = 0.0617$ (Frank et al., 2013).

2-Choose critical value C: Given Hong & Raudenbush (2005) reported that kindergarten retention had a significant negative effect on reading achievement, we decide to choose C as -1.96, or equivalently $\beta_W^{\#} = -1.96 se(\hat{\beta}_W^{id})$.

3-Obtain the relationship between the PIV and the mean counterfactual outcomes: After the observed sample statistics and the critical value are plugged into the probit model (9), the PIV is the probit function of the mean counterfactual reading score for the retained students had they been promoted instead (i.e., $\bar{Y}_c^{un}$) and the mean counterfactual reading score for the promoted students had they been retained instead (i.e., $\bar{Y}_t^{un}$) as follows:

$$probit(PIV) = -1.96 - \frac{109.25*(0.9383\bar{Y}_t^{un} - 0.0617\bar{Y}_c^{un} - 40.69)}{\sqrt{564.18 + 0.116*[(\bar{Y}_t^{un} - 36.77)^2 + (\bar{Y}_c^{un} - 45.78)^2] + (0.9383\bar{Y}_t^{un} - 0.0617\bar{Y}_c^{un} - 40.69)^2}} \quad (10)$$

4-State belief about the mean counterfactual outcomes: This step requires one to form and bound belief about the two mean counterfactual outcomes. To illustrate this procedure, we form two different beliefs about them.

4.1-The first belief: Given the inference of Hong & Raudenbush (2005) mostly addressed $\bar{Y}_c^{un}$, i.e., the mean counterfactual reading score of the retained students, we decide to bound $\bar{Y}_t^{un}$ and assume $\bar{Y}_c^{un} = 45.2$. We choose this value because it is the grand sample mean so that $\bar{Y}_t^{un} - \bar{Y}_c^{un}$ measures the degree to which the counterfactual reading scores deviate from the null hypothesis: $\beta_W = 0$. The probit model (10) is thus simplified as:

$$probit(PIV) = -1.96 - \frac{102.51\bar{Y}_t^{un} - 4750.19}{\sqrt{564.22 + 0.116(\bar{Y}_t^{un} - 36.77)^2 + (0.9383\bar{Y}_t^{un} - 43.48)^2}} \quad (11)$$

In this case, one may ask "what would be the mean reading score of the promoted students had they been retained instead (i.e., $\bar{Y}_t^{un}$)?" when the mean reading score of the retained students had they been promoted (i.e., $\bar{Y}_c^{un}$) is assumed to be 45.2. This might be answered by reflecting on the counterfactual outcomes based on belief about the average retention effects for the retained students and for the promoted students, identified by $\bar{Y}_t^{ob} - \bar{Y}_c^{un}$ and $\bar{Y}_t^{un} - \bar{Y}_c^{ob}$ respectively. For example, given the average retention effect for the retained students is significantly negative (36.77-45.2 = -8.43), it is reasonable to think the average retention effect for the promoted students should be at least negative as well (Jimerson, 2001; Lorence et al., 2002; Burkam et al., 2007). This leads to the upper bound for $\bar{Y}_t^{un}$ as 45.78.

4.2-The second belief: As in the first belief, we believe that kindergarten retention should have negative impact on reading achievement for both retained and promoted students. In addition, we believe that the original estimate of average retention effect for the retained students, which was -9, was overestimated such that the inference could be invalidated. Therefore, a plausible region for the mean counterfactual outcomes is defined as $\bar{Y}_t^{un} \leq 45.78$ and $36.77 \leq \bar{Y}_c^{un} \leq 45.78$. Built on this plausible region, our second belief assumes that the mean counterfactual reading score for the promoted students could not exceed the grand sample mean had they been retained instead, i.e., $\bar{Y}_t^{un} \leq 45.2$. The plausible region is illustrated by figure 2.

5-Bound the PIV: For our first belief, the lower bound of the PIV is 0.92. This means the chance that the inference of Hong & Raudenbush (2005) is robust for internal validity is at least 92% if the mean reading score of the promoted students had they been retained instead is believed to be at most 45.78 and the mean reading score of the retained students had they been promoted instead is believed to be 45.2. For our second belief, the lower bound of the PIV is 0.936. This suggests the chance that Hong & Raudenbush's inference is robust for internal validity is at least 93.6% if the mean reading score of the promoted students had they been retained instead is believed to be at most 45.2 and the mean reading score of the retained students had they been promoted instead is believed to be at least 36.77.

6-Conclusion: To reach a clear conclusion about internal validity, one can use a threshold about the PIV such that an inference is deemed robust for internal validity whenever the PIV exceeds this threshold. Since the PIV is the statistical power of retesting the null hypothesis: $\delta = 0$ based on the ideal sample, we use PIV = 0.8 as the threshold which is often used for strong statistical power (Cohen 1988, 1992). Therefore, the two beliefs we formed in the fourth step would lead to the conclusion that Hong & Raudenbush's inference is robust for internal validity. We caution readers that this conclusion might not hold if one has a different belief and/or a different threshold for the PIV.

Nevertheless, a decision about the internal validity of Hong & Raudenbush (2005) is less clear for the entire plausible region, as shown by figure 2. Still, there are two key observations to be noted: First, in general the PIV is more sensitive to $\bar{Y}_t^{un}$ than $\bar{Y}_c^{un}$, since most observed students actually got promoted. This suggests the inference of Hong & Raudenbush is likely to be robust as long as kindergarten retention is believed to have stronger-than-minimal negative impact on the promoted students. Second, even if the kindergarten retention has minimal negative impact on the promoted students ( $45.2 < \bar{Y}_t^{un} < 45.78$ ), the inference of Hong and Raudenbush (2005) would still be robust for internal validity as long as the average retention effect for the retained students was just slightly overestimated. For example, the lower bound of the PIV is 0.795 when the average retention effect for the retained students was believed to be at least -7 ( $\bar{Y}_c^{un} \geq 43.77$ ), which is 22% smaller in size than the original estimate. However, the inference of Hong and Raudenbush (2005) would be questionable if the impact of kindergarten retention on the promoted students is minimal and it was severely over-estimated for the retained students, hinting the existence of extremely strong confounder(s).

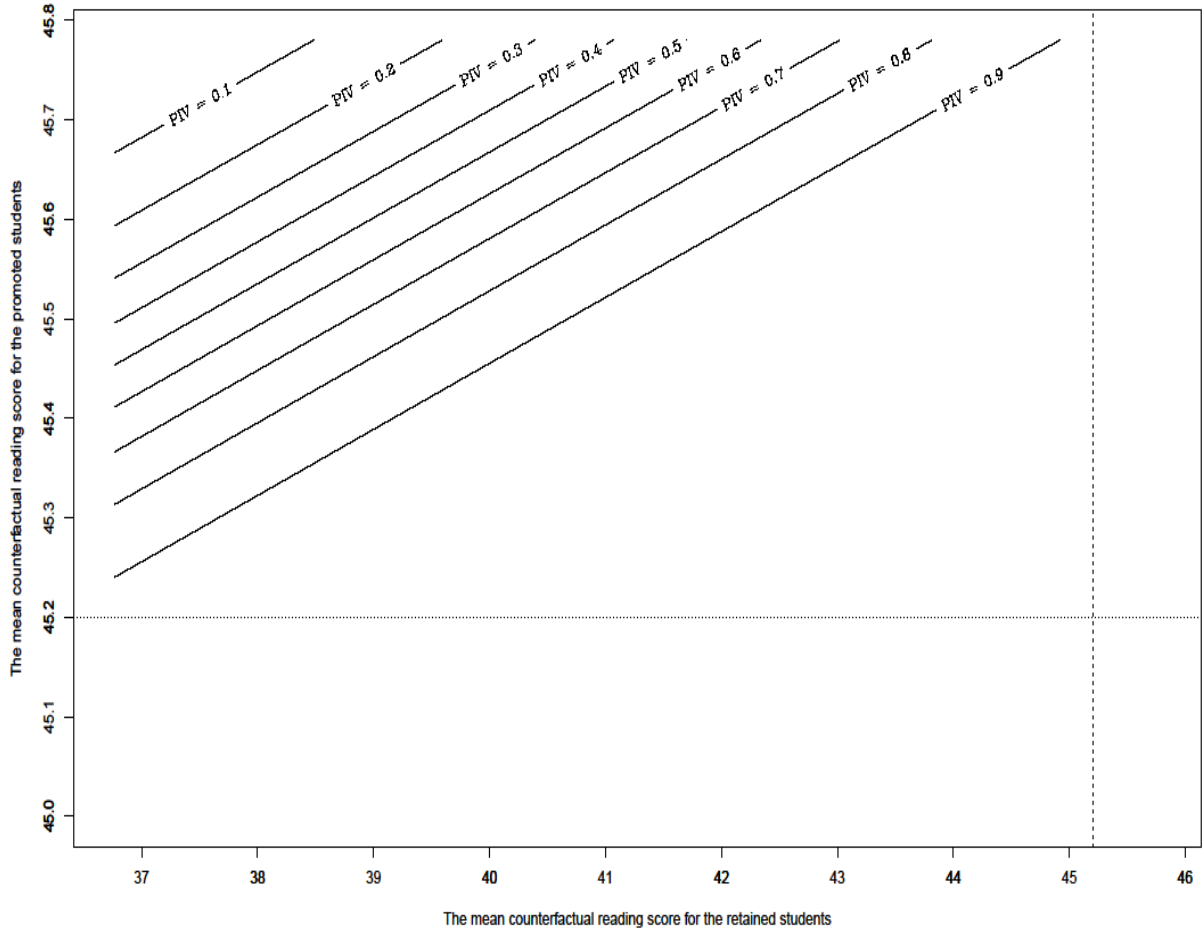

*Figure 2:* The contour plot of the PIV in the plausible region with the x axis representing $\bar{Y}_c^{un}$ and the y axis representing $\bar{Y}_t^{un}$. The plausible region is defined based on the belief that the average retention effect for the promoted students should not be positive and the average retention effect for the retained students was overestimated, which means both $\bar{Y}_t^{un}$ and $\bar{Y}_c^{un}$ are smaller than 45.78. The vertical dashed line corresponds to the first belief where $\bar{Y}_c^{un} = 45.2$ and the horizontal dashed line corresponds to the second belief where $\bar{Y}_t^{un} < 45.2$.

By definition, the PIV equals the statistical power of testing the null hypothesis: $\beta_W = 0$ versus the alternative hypothesis: $\beta_W = \hat{\beta}_W^{id}$ ($\hat{\beta}_W^{id} \neq 0$), had the counterfactual data became available. This is illustrated by figure 3 created by assuming $\bar{Y}_c^{un} = 45.2$. It is clear that, as $\bar{Y}_t^{un}$ decreases, the distribution of beta coefficient for kindergarten retention in the ideal sample shifts to the left, which increases the PIV. Figure 3 demonstrates how the PIV is equivalent to the statistical power of retesting the null hypothesis as if the counterfactual data were available. Interpretatively, assessing internal validity through PIV can be thought of as power analysis for the ideal sample.

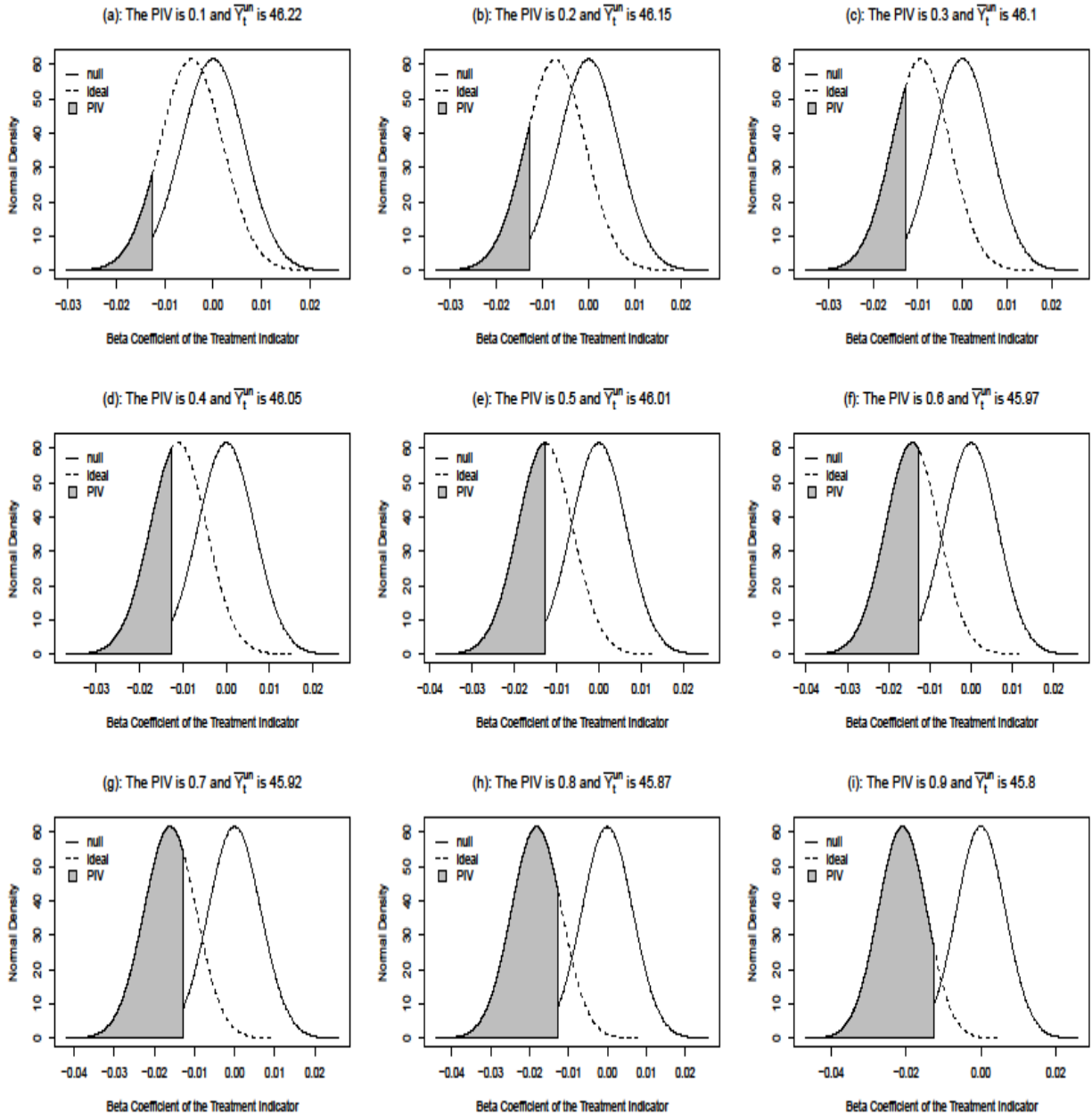

*Figure 3*: The relationship between the PIV and retesting hypothesis in the ideal sample for Hong and Raudenbush (2005), assuming $\bar{Y}_c^{un} = 45.2$. The solid curve represents the null hypothesis: $\beta_W = 0$ and the dashed curve represents the alternative hypothesis: $\beta_W = \hat{\beta}_W^{id}$. The grey shaded area symbolizes the PIV of Hong & Raudenbush.

**6-Literature review on similar methods**

Literature review on sensitivity analysis: Sensitivity analysis (Rosenbaum and Rubin, 1983b; Rosenbaum, 1986, 1987, 1991, 2002, 2010) evaluates the impact of a missing confounder on regression estimates and nonparametric tests by conceptually connecting violation of the unconfoundedness assumption to violation of random assignment in matched pairs. Therefore, it informs the internal validity of a matching design. Other literature on sensitivity analysis uses similar approaches for potential failure of the unconfoundedness assumption (Copas and Li, 1997; Lin et al., 1998; Robins et al., 2000; VanderWeele, 2008; Hosman et al., 2010; Masten and Poirier, 2018). The PIV shares the objective of checking the sensitivity of results to potential failure of the unconfoundedness assumption, but it is not limited to a single type of design (such as matching) or model (such as regression). In fact, the PIV can be applied to any design deemed appropriate for observational studies.

Literature review on Bayesian sensitivity analysis: Bayesian sensitivity analysis (BSA) (McCandless et al. 2007, 2012; McCandless and Gustafson, 2017) parameterizes the models for the outcome and the unmeasured confounder carefully to identify the key confounding parameters and treat the unmeasured confounder as missing data under a Bayesian framework. BSA has two main advantages: First, data augmentation allows one to repeatedly impute missing values for the unobserved confounder. This allows one to take the unobserved confounder into account when generating the joint posterior distribution of the confounding and treatment effect parameters. Additionally, BSA offers modeling flexibility through specification of the prior. Compared to BSA, analysis using the PIV is more approachable and interpretable as it does not require a complicated MCMC algorithm and is built on the common NHST framework.

Literature review on robustness indices of causal inferences: Robustness indices of causal inferences (Frank, 2000; Frank et al., 2013) quantify the strength of internal validity in terms of the impact of an unmeasured confounding variable or the proportion of observed cases can be replaced by null cases that an inference can withstand. The PIV is inherently connected to both papers as it shares the decision rules and the missing data perspective with Frank et al. (2013) and the focus on the relationship between the estimate of average treatment effect and the NHST with Frank (2000). The PIV is different from the robustness indices because it needs a bounded belief about the mean counterfactual outcomes as input and it is a probabilistic index comparable to statistical power.

Literature review on bounding treatment effects: Bounding treatment effects is motivated by the non-identification issue of an average treatment effect due to counterfactual outcomes (Altonji et al., 2005, Manski, 1990, 1995; Manski and Nagin, 1998). Bounds are obtained by imposing assumptions on the counterfactuals and can be tightened by making the assumption(s) more informative. Both the PIV and Manski's bounds of treatment effect consider situations when the unconfoundedness assumption is implausible so that stronger assumptions about counterfactual outcomes are needed to inform inferences. Different from the PIV, Manski's bounds are not built on NHST and parametric assumption. Rather, the bounds offer insights about the worth of a causal inference through exploring loss-based alternatives rooted in the context of program evaluation. Furthermore, Manski's bounds leverage non-linear relationships to determine constraints on parameter values, whereas the PIV is built on the classical linear regression model (CLRM) and quantifies the likelihood an inference would hold under CLRM.

Literature review on replication probabilities: Replication probabilities have been promoted for two main reasons: First, they purpose safeguarding readers from misguidance and misinterpretation of the p-values. Second, they are used to accentuate that the true scientific significance concerning replicability rather than statistical significance (Greenwald et al, 1996; Posavac, 2002; Shao and Chow, 2002; Killeen, 2005; Boos and Stefanski, 2011). The PIV is actually the probability of replicating a significant result in an observational study for the ideal sample, and it is similar to $p_{rep}$ (Killeen, 2005; Iverson et al., 2010) which is the probability of obtaining an effect with the same sign as the observed one. Different from $p_{rep}$ and other replication probabilities, the PIV takes counterfactual outcomes into consideration and therefore it is not a function of the p-value. Therefore, it does not inherit any of the limitations associated with the p-value as most proposed replication probabilities do (Doros and Geier, 2005).

**7-Discussion**

Focusing on the beta coefficient of treatment indicator, we began by defining the unobserved sample as the collection of the counterfactual data and the ideal sample as the combination of the unobserved sample and the observed sample. The assessment of internal validity should be based on the ideal sample and for null hypothesis statistical testing (NHST) this means one should test the null hypothesis (versus the alternative) in the ideal sample and check if the result is consistent with the testing result based on the observed sample. The probability of a causal inference is robust for internal validity, i.e., the PIV, is thus defined as the probability of rejecting the same null hypothesis again in the ideal sample given it has been rejected in the observed sample. Internal validity is evaluated by bounding the PIV of an inference based on a bounded belief about the mean counterfactual outcomes.

The scholarly significance of this study manifests in three aspects: First, it prompts researchers to conceptualize the counterfactual outcomes and form bounded beliefs about them. This will foster critical thinking as well as scientific discourse about internal validity since people can use the PIV to understand under what circumstances and to what degree internal validity will be robust. Second, the PIV can be interpreted as the statistical power of testing the hypothesis $H_0 : \beta_W = 0$ versus $H_a : \beta_W = \hat{\beta}_W^{id}$ in the ideal sample. Therefore, the PIV is pragmatic as it quantifies the impact of counterfactual outcomes (and thus internal validity) on decision-making. Third, the PIV is built on a simple Bayesian framework with intuitive missing data interpretations. Future work should focus on extending the PIV in two ways: First, future work should modify the current framework for nonlinear models and cases where CLRM is implausible such as longitudinal studies. Second, built on the framework for how counterfactual outcomes affect the NHST through the PIV, future work needs to investigate further why counterfactual outcomes change, which may due to an omitted confounder, the violation of stable unit treatment value assumption (SUTVA) or measurement error.

Shadish, W. R., Cook, T. D., & Campbell, D. T. (2002). *Experimental and quasi-experimental designs for generalized causal inference*. New York, NY: Houghton Mifflin.

Shao, J., & Chow, S. C. (2002). Reproducibility probability in clinical trials. *Statistics in Medicine*, *21*(12), 1727-1742.

Sobel, M. E. (1996). An introduction to causal inference. *Sociological Methods & Research* 24(3): 353-379.

Tingle, L. R., Schoeneberger, J., & Algozzine, B. (2012). Does grade retention make a difference? *The Clearing House: A Journal of Educational Strategies, Issues and Ideas,* 85(5), 179-185.

VanderWeele, T. J. (2008). Sensitivity analysis: distributional assumptions and confounding assumptions. *Biometrics* 64(2): 645-649.30

# Appendix

## Proofs of Theorem 1 and Theorem 2

Proof of theorem 1:

Our goal is to derive the formula for least square estimate of regression coefficient for W (i.e., $\hat{\beta}_W^{id}$) based on the ideal sample and the variance of $\hat{\beta}_W^{id}$, since they are the mean and the variance of the distribution of $\beta_W$ conditional on the ideal sample. First, one needs to define the following ordered data matrices for the ideal sample:

$$
\begin{aligned}
\mathbf{D} &= [\mathbf{Y}_{(n^{un}+n^{ob})\times 1}, \mathbf{X}_{(n^{un}+n^{ob})\times(p+2)}] = [\mathbf{Y}_{2n^{ob}\times 1}, \mathbf{X}_{2n^{ob}\times(p+2)}] \\
\mathbf{X} &= [\mathbf{1}_{(n^{un}+n^{ob})\times 1}, \mathbf{V}_{(n^{un}+n^{ob})\times(p+1)}] = [\mathbf{1}_{2n^{ob}\times 1}, \mathbf{V}_{2n^{ob}\times(p+1)}] \\
\mathbf{V} &= [\mathbf{Z}_{(n^{un}+n^{ob})\times p}, \mathbf{W}_{(n^{un}+n^{ob})\times 1}] = \begin{bmatrix} \mathbf{Z}^{ob}_{n^{ob}\times p}, \mathbf{1}_{n^{ob}\times 1} - \mathbf{W}^{ob}_{n^{ob}\times 1} \\ \mathbf{Z}^{ob}_{n^{ob}\times p}, \mathbf{W}^{ob}_{n^{ob}\times 1} \end{bmatrix} \\
\mathbf{Z} &= [\mathbf{Z_1}, \mathbf{Z_2}, ..., \mathbf{Z_p}]_{(n^{un}+n^{ob})\times p} = [\mathbf{Z_1}, \mathbf{Z_2}, ..., \mathbf{Z_p}]_{2n^{ob}\times p}
\end{aligned}
\quad (A1)
$$

and the following ordered mean vectors:

$$
\begin{aligned}
\overline{\mathbf{V}}^{\mathbf{id}} &= [\overline{\mathbf{Z}}^{\mathbf{id}}, \overline{W}^{id}]_{1\times(p+1)} \\
\overline{\mathbf{Z}}^{\mathbf{id}} &= [\overline{Z}_1^{id}, \overline{Z}_2^{id}, \cdots, \overline{Z}_p^{id}]_{1\times p}
\end{aligned}
\quad (A2)
$$

The matrix $\mathbf{X}^T\mathbf{X}$ for the ideal sample could then be molded as the following block matrix:

$$
\mathbf{X}^T\mathbf{X} = \begin{pmatrix} n^{un}+n^{ob} & (n^{un}+n^{ob})\overline{\mathbf{V}}^{\mathbf{id}} \\ (n^{un}+n^{ob})(\overline{\mathbf{V}}^{\mathbf{id}})^{\mathbf{T}} & \mathbf{V}^{\mathbf{T}}\mathbf{V} \end{pmatrix}
\quad (A3)
$$

The inverse of $\mathbf{X}^T\mathbf{X}$ can be shown to have the following form:

$$(\mathbf{X}^T\mathbf{X})^{-1} = \begin{pmatrix} \dfrac{1}{n^{un}+n^{ob}} + \bar{\mathbf{V}}^{id}\dfrac{1}{n^{un}+n^{ob}}(\mathbf{S}_{VV}^{id})^{-1}(\bar{\mathbf{V}}^{id})^T & -\dfrac{1}{n^{un}+n^{ob}}\bar{\mathbf{V}}^{id}(\mathbf{S}_{VV}^{id})^{-1} \\ -\dfrac{1}{n^{un}+n^{ob}}(\mathbf{S}_{VV}^{id})^{-1}(\bar{\mathbf{V}}^{id})^T & \dfrac{1}{n^{un}+n^{ob}}(\mathbf{S}_{VV}^{id})^{-1} \end{pmatrix}$$

(A4)

It should be clear now that, to determine the definite form of $(\mathbf{X}^T\mathbf{X})^{-1}$ I need to find out what $(\mathbf{S}_{VV}^{id})^{-1}$ is. As a variance-covariance matrix for the vector of predictors V, $\mathbf{S}_{VV}^{id}$ can be expressed as the block matrix whose elements is formalized below:

$$\mathbf{S}_{VV}^{id} = \begin{pmatrix} \mathbf{S}_{ZZ}^{id} & \mathbf{S}_{ZW}^{id} \\ \mathbf{S}_{WZ}^{id} & \hat{\sigma}_{WW}^{id} \end{pmatrix}_{(p+1)\times(p+1)}$$

(A5)

where:

$$\mathbf{S}_{ZZ}^{id} = \begin{pmatrix} \hat{\sigma}_{Z_1Z_1}^{id} & \cdots & \hat{\sigma}_{Z_1Z_p}^{id} \\ \vdots & \ddots & \vdots \\ \hat{\sigma}_{Z_pZ_1}^{id} & \cdots & \hat{\sigma}_{Z_pZ_p}^{id} \end{pmatrix}_{p\times p}$$

$$\mathbf{S}_{ZW}^{id} = \begin{pmatrix} \hat{\sigma}_{Z_1W}^{id} \\ \vdots \\ \hat{\sigma}_{Z_pW}^{id} \end{pmatrix}_{p\times 1}$$

$$\mathbf{S}_{WZ}^{id} = \begin{pmatrix} \hat{\sigma}_{Z_1W}^{id} & \cdots & \hat{\sigma}_{Z_pW}^{id} \end{pmatrix}_{1\times p}$$

(A6)

Furthermore, we define the following covariance vector:

$$\mathbf{S}_{\mathbf{ZY}}^{\mathbf{id}} = \begin{pmatrix} \hat{\sigma}_{Z_1 Y}^{id} \\ \vdots \\ \hat{\sigma}_{Z_p Y}^{id} \end{pmatrix}_{p \times 1} \tag{A7}$$

All aforementioned sample covariances and variances are supposed to be computed according to the following formula:

$$\hat{\sigma}_{xy} = \frac{1}{n} \sum_{i=1}^{n} (x_i - \bar{x})(y_i - \bar{y})$$
$$\hat{\sigma}_{xx} = \frac{1}{n} \sum_{i=1}^{n} (x_i - \bar{x})^2 \tag{A8}$$

for any variable x or y and any sample size n in this context.

Consequently, the inverse of $\mathbf{S}_{\mathbf{VV}}^{\mathbf{id}}$ can be formulated here:

$$(\mathbf{S}_{\mathbf{VV}}^{\mathbf{id}})^{-1} =$$
$$\begin{bmatrix} (\mathbf{S}_{\mathbf{ZZ}}^{\mathbf{id}})^{-1} + (\mathbf{S}_{\mathbf{ZZ}}^{\mathbf{id}})^{-1}\mathbf{S}_{\mathbf{ZW}}^{\mathbf{id}}(\hat{\sigma}_{WW}^{id} - \mathbf{S}_{\mathbf{WZ}}^{\mathbf{id}}(\mathbf{S}_{\mathbf{ZZ}}^{\mathbf{id}})^{-1}\mathbf{S}_{\mathbf{ZW}}^{\mathbf{id}})^{-1}\mathbf{S}_{\mathbf{WZ}}^{\mathbf{id}}(\mathbf{S}_{\mathbf{ZZ}}^{\mathbf{id}})^{-1} & -(\mathbf{S}_{\mathbf{ZZ}}^{\mathbf{id}})^{-1}\mathbf{S}_{\mathbf{ZW}}^{\mathbf{id}}(\hat{\sigma}_{WW}^{id} - \mathbf{S}_{\mathbf{WZ}}^{\mathbf{id}}(\mathbf{S}_{\mathbf{ZZ}}^{\mathbf{id}})^{-1}\mathbf{S}_{\mathbf{ZW}}^{\mathbf{id}})^{-1} \\ -(\hat{\sigma}_{WW}^{id} - \mathbf{S}_{\mathbf{WZ}}^{\mathbf{id}}(\mathbf{S}_{\mathbf{ZZ}}^{\mathbf{id}})^{-1}\mathbf{S}_{\mathbf{ZW}}^{\mathbf{id}})^{-1}\mathbf{S}_{\mathbf{WZ}}^{\mathbf{id}}(\mathbf{S}_{\mathbf{ZZ}}^{\mathbf{id}})^{-1} & (\hat{\sigma}_{WW}^{id} - \mathbf{S}_{\mathbf{WZ}}^{\mathbf{id}}(\mathbf{S}_{\mathbf{ZZ}}^{\mathbf{id}})^{-1}\mathbf{S}_{\mathbf{ZW}}^{\mathbf{id}})^{-1} \end{bmatrix}$$

(A9)

Plugging the above matrix of $(\mathbf{S}_{\mathbf{VV}}^{\mathbf{id}})^{-1}$ into the block matrix of $(\mathbf{X}^\mathbf{T}\mathbf{X})^{-1}$ will give us the complete definite form of matrix $(\mathbf{X}^\mathbf{T}\mathbf{X})^{-1}$, whose elements are all ideal sample statistics such as ideal sample variances, ideal sample covariances and ideal sample means. To isolate the estimated regression coefficient for W, I only need to use the elements in the last row of $(\mathbf{X}^\mathbf{T}\mathbf{X})^{-1}$, which are provide next:

$$(\mathbf{X^TX})^{-1}{}_{(p+2)1} = \frac{1}{n^{un}+n^{ob}}[(\hat{\sigma}_{WW}^{id} - \mathbf{S}_{WZ}^{id}(\mathbf{S}_{ZZ}^{id})^{-1}\mathbf{S}_{ZW}^{id})^{-1}\mathbf{S}_{WZ}^{id}(\mathbf{S}_{ZZ}^{id})^{-1}(\mathbf{\bar{Z}}^{id})^{T} - \bar{W}^{id}(\hat{\sigma}_{WW}^{id} - \mathbf{S}_{WZ}^{id}(\mathbf{S}_{ZZ}^{id})^{-1}\mathbf{S}_{ZW}^{id})^{-1}]$$

$$[(\mathbf{X^TX})^{-1}{}_{(p+2)2},\cdots,(\mathbf{X^TX})^{-1}{}_{(p+2)(p+1)}] = -\frac{1}{n^{un}+n^{ob}}(\hat{\sigma}_{WW}^{id} - \mathbf{S}_{WZ}^{id}(\mathbf{S}_{ZZ}^{id})^{-1}\mathbf{S}_{ZW}^{id})^{-1}\mathbf{S}_{WZ}^{id}(\mathbf{S}_{ZZ}^{id})^{-1}$$

$$(\mathbf{X^TX})^{-1}{}_{(p+2)(p+2)} = \frac{1}{n^{un}+n^{ob}}(\hat{\sigma}_{WW}^{id} - \mathbf{S}_{WZ}^{id}(\mathbf{S}_{ZZ}^{id})^{-1}\mathbf{S}_{ZW}^{id})^{-1}$$

(A10)

Because the estimated regression coefficient for W is the last element of $(\mathbf{X^TX})^{-1}\mathbf{X^TY}$ which is the dot product between the last row of $(\mathbf{X^TX})^{-1}$ and $\mathbf{X^TY}$, the expression of $\mathbf{X^TY}$ is also needed here:

$$\mathbf{X^TY} = \begin{bmatrix} (n^{un}+n^{ob})\bar{Y}^{id} \\ \mathbf{Z^TY} \\ \mathbf{W^TY} \end{bmatrix} \quad (A11)$$

where:

$$\mathbf{Z^TY} = (n^{un}+n^{ob})\mathbf{S}_{ZY}^{id} + (n^{un}+n^{ob})\bar{Y}^{id}(\mathbf{\bar{Z}}^{id})^{T}$$
$$\mathbf{W^TY} = (n^{un}+n^{ob})\hat{\sigma}_{WY}^{id} + (n^{un}+n^{ob})\bar{W}^{id}\bar{Y}^{id} \quad (A12)$$

Now one can calculate the estimated regression coefficient for W as the dot product between the last row of $(\mathbf{X^TX})^{-1}$ and the vector $\mathbf{X^TY}$. The result is presented below:

$$\hat{\beta}_W^{id} = \frac{\hat{\sigma}_{WY}^{id} - \mathbf{S}_{WZ}^{id}(\mathbf{S}_{ZZ}^{id})^{-1}\mathbf{S}_{ZY}^{id}}{\hat{\sigma}_{WW}^{id} - \mathbf{S}_{WZ}^{id}(\mathbf{S}_{ZZ}^{id})^{-1}\mathbf{S}_{ZW}^{id}} \quad (A13)$$

The variance of $\hat{\beta}_W^{id}$ should be straightforward: it is just the product of the known residual variance $\sigma^2$ and the element in the (p+2)$^{th}$ row and the (p+2)$^{th}$ column of $(\mathbf{X}^T\mathbf{X})^{-1}$:

$$Var(\hat{\beta}_W^{id}) = \frac{\sigma^2}{n^{un} + n^{ob}} (\hat{\sigma}_{WW}^{id} - \mathbf{S}_{WZ}^{id}(\mathbf{S}_{ZZ}^{id})^{-1}\mathbf{S}_{ZW}^{id})^{-1} \tag{A14}$$

Taken together, the distribution of $\beta_W$ given the ideal sample for the simple estimator is:

$$\beta_W \mid \mathbf{D}^{id} \sim N(\frac{\hat{\sigma}_{WY}^{id} - \mathbf{S}_{WZ}^{id}(\mathbf{S}_{ZZ}^{id})^{-1}\mathbf{S}_{ZY}^{id}}{\hat{\sigma}_{WW}^{id} - \mathbf{S}_{WZ}^{id}(\mathbf{S}_{ZZ}^{id})^{-1}\mathbf{S}_{ZW}^{id}}, \frac{\sigma^2}{n^{un} + n^{ob}}(\hat{\sigma}_{WW}^{id} - \mathbf{S}_{WZ}^{id}(\mathbf{S}_{ZZ}^{id})^{-1}\mathbf{S}_{ZW}^{id})^{-1})$$

(A15)

(A15) can be greatly simplified in observational studies, given $\mathbf{S}_{WZ}^{id} = \mathbf{0}$ and $n^{un} = n^{ob}$:

$$\beta_W \mid \mathbf{D}^{id} \sim N(\frac{\hat{\sigma}_{WY}^{id}}{\hat{\sigma}_{WW}^{id}}, \frac{\sigma^2}{2n^{ob}}(\hat{\sigma}_{WW}^{id})^{-1}) \tag{A16}$$

If $\beta_W$ is the standardized regression coefficient, (A16) could be further simplified as follows:

$$\beta_W \mid \mathbf{D}^{id} \sim N(r_{wy}^{id}, \frac{1-R^2}{2n^{ob}}) \tag{A17}$$

The expression of $r_{wy}^{id}$ would be given by the formulae below, according to Cohen et al. (2003):

$$r_{wy}^{id} = \left(\frac{\bar{Y}_t^{id} - \bar{Y}_c^{id}}{\sigma_Y^{id}}\right) 0.5 \tag{A18}$$

The expression for $\sigma_Y^{id}$ is derived as:

$$\sigma_Y^{id} = \sqrt{0.5[\hat{\sigma}_t^2 + \pi(1-\pi)(\bar{Y}_t^{un} - \bar{Y}_t^{ob})^2] + 0.5[\hat{\sigma}_c^2 + \pi(1-\pi)(\bar{Y}_c^{un} - \bar{Y}_c^{ob})^2] + 0.25(\bar{Y}_t^{id} - \bar{Y}_c^{id})^2}$$

$$= \sqrt{0.5\hat{\sigma}_t^2 + 0.5\pi(1-\pi)[(\bar{Y}_t^{un} - \bar{Y}_t^{ob})^2 + (\bar{Y}_c^{un} - \bar{Y}_c^{ob})^2] + 0.5\hat{\sigma}_c^2 + 0.25(\bar{Y}_t^{id} - \bar{Y}_c^{id})^2}$$

(A19)

Combining (A18) and (A19), $r_{wy}^{id}$ is derived as follows:

$$r_{wy}^{id} = \frac{\bar{Y}_t^{id} - \bar{Y}_c^{id}}{\sqrt{2\hat{\sigma}_t^2 + 2\pi(1-\pi)[(\bar{Y}_t^{un} - \bar{Y}_t^{ob})^2 + (\bar{Y}_c^{un} - \bar{Y}_c^{ob})^2] + 2\hat{\sigma}_c^2 + (\bar{Y}_t^{id} - \bar{Y}_c^{id})^2}} \quad (A20)$$

where:

$$\bar{Y}_t^{id} = (1-\pi)\bar{Y}_t^{un} + \pi\bar{Y}_t^{ob}$$
$$\bar{Y}_c^{id} = \pi\bar{Y}_c^{un} + (1-\pi)\bar{Y}_c^{ob} \quad (A21)$$

Finally, plugging the expression (A20) and (A21) into the distribution (A17) will lead to the distribution of true average effect proposed in theorem 1.

The next step is to prove theorem 1 under Bayesian framework and show the posterior distribution is identical to the distribution we derived above when the prior distribution is thought to be based on the unobserved sample:

If the prior and likelihood for regression coefficients $\boldsymbol{\beta}$ are set as follows:

$$\boldsymbol{\beta} \sim N(((\mathbf{X^{un}})^T\mathbf{X^{un}})^{-1}(\mathbf{X^{un}})^T\mathbf{Y^{un}}, \sigma^2((\mathbf{X^{un}})^T\mathbf{X^{un}})^{-1})$$
$$Y_i | \boldsymbol{\beta}, \mathbf{X_i^{ob}} \sim N(\mathbf{X_i^{ob}}\boldsymbol{\beta}, \sigma^2) \quad (A22)$$

with the following defined data matrices:

$$\mathbf{X^{ob}} = [\mathbf{1}_{n^{ob} \times 1}, \mathbf{W}^{ob}_{n^{ob} \times 1}, \mathbf{Z}^{ob}_{n^{ob} \times p}]$$
$$\mathbf{X^{un}} = [\mathbf{1}_{n^{ob} \times 1}, \mathbf{1}_{n^{ob} \times 1} - \mathbf{W}^{ob}_{n^{ob} \times 1}, \mathbf{Z}^{ob}_{n^{ob} \times p}] \quad (A23)$$

That is, for the i[th] subject, the unobserved covariate vector will be the same as its observed covariate vector except that its treatment status is changed.

Then the posterior of $\boldsymbol{\beta}$ is:

$$\boldsymbol{\beta} \mid \mathbf{D^{ob}} \sim N(\boldsymbol{\theta_\beta}, \boldsymbol{\Phi_\beta}) \quad (A24)$$

where:

$$\boldsymbol{\theta_\beta} = ((\mathbf{X^{un}})^T \mathbf{X^{un}} + (\mathbf{X^{ob}})^T \mathbf{X^{ob}})^{-1}((\mathbf{X^{un}})^T \mathbf{Y^{un}} + (\mathbf{X^{ob}})^T \mathbf{Y^{ob}})$$
$$\boldsymbol{\Phi_\beta} = \sigma^2 ((\mathbf{X^{un}})^T \mathbf{X^{un}} + (\mathbf{X^{ob}})^T \mathbf{X^{ob}})^{-1} \quad (A25)$$

Moreover, the following equations hold for the ideal sample:

$$(\mathbf{X^{id}})^T \mathbf{X^{id}} = (\mathbf{X^{un}})^T \mathbf{X^{un}} + (\mathbf{X^{ob}})^T \mathbf{X^{ob}}$$
$$(\mathbf{X^{id}})^T \mathbf{Y^{id}} = (\mathbf{X^{un}})^T \mathbf{Y^{un}} + (\mathbf{X^{ob}})^T \mathbf{Y^{ob}} \quad (A26)$$

Combining (A25) and (A26), it is straightforward that the posterior distribution of $\boldsymbol{\beta}$ is identical to the distribution of $\boldsymbol{\beta}$ given the ideal sample, which is normally distributed with the mean as $((\mathbf{X^{id}})^T \mathbf{X^{id}})^{-1}((\mathbf{X^{id}})^T \mathbf{Y^{id}})$ and the variance as $\sigma^2((\mathbf{X^{id}})^T \mathbf{X^{id}})^{-1}$. The posterior distribution of $\beta_W$, which is just the marginal distribution of $\boldsymbol{\beta}$ in (A24), should then be identical to (A16) which is the marginal distribution of $\beta_W$ in the distribution of $\boldsymbol{\beta}$ given the ideal sample. To conclude, if the prior distribution is thought of as the distribution of the regression coefficients $\boldsymbol{\beta}$ based on the unobserved sample, the posterior distribution can then be thought of as the distribution of regression coefficients $\boldsymbol{\beta}$ conditional on the ideal sample.

Proof of theorem 2:

Based on theorem 1, the PIV is shown to be a function of the ideal sample statistics. Specifically, when a significant positive effect has been concluded, the PIV is written as follows:

$$PIV = P(\beta_W > \beta_W^\# \mid \mathbf{D^{id}}) =$$

$$P\left( \frac{\beta - r_{wy}^{id}}{\sqrt{\dfrac{1-R^2}{2n^{ob}}}} > \frac{\beta_W^\# - r_{wy}^{id}}{\sqrt{\dfrac{1-R^2}{2n^{ob}}}} \mid \mathbf{D^{id}} \right)$$

$$= 1 - \Phi\left( \frac{\beta_W^\# - r_{wy}^{id}}{\sqrt{\dfrac{1-R^2}{2n^{ob}}}} \right) \quad \text{(A27)}$$

$$= \Phi\left( \frac{\sqrt{2n^{ob}}}{\sqrt{1-R^2}} [r_{wy}^{id} - \beta_W^\#] \right)$$

From (A27), the probit model for the PIV can be shown to be identical to (5).

Likewise, when a significant negative effect has been concluded, the PIV is expressed as follows:

$$PIV = P(\beta_W < \beta_W^{\#} \mid \mathbf{D^{id}}) =$$

$$P\left( \frac{\beta - r_{wy}^{id}}{\sqrt{\frac{1-R^2}{2n^{ob}}}} < \frac{\beta_W^{\#} - r_{wy}^{id}}{\sqrt{\frac{1-R^2}{2n^{ob}}}} \,\Bigg|\, \mathbf{D^{id}} \right)$$

$$= \Phi\left( \frac{\beta_W^{\#} - r_{wy}^{id}}{\sqrt{\frac{1-R^2}{2n^{ob}}}} \right) \qquad (A28)$$

$$= \Phi\left( \frac{\sqrt{2n^{ob}}}{\sqrt{1-R^2}} [\beta_W^{\#} - r_{wy}^{id}] \right)$$

From (A28), the probit model for the PIV can be shown to be identical to (6).